\documentclass{aa}  
\usepackage[normalem]{ulem}
\usepackage{graphicx}
\usepackage{comment}
\usepackage[acronym]{glossaries}
\usepackage{booktabs} 
\usepackage{xcolor}
\usepackage{url}
\usepackage[pdfencoding=auto,psdextra]{hyperref}
\hypersetup{
    colorlinks=true,
    linkcolor=blue,
    filecolor=magenta,      
    urlcolor=blue,
    citecolor=blue
}

\newacronym{bh}{BH}{}
\newacronym{emris}{EMRIs}{}
\newacronym{emri}{EMRI}{}
\newacronym{mbh}{MBH}{}
\newacronym{gw}{GW}{}
\newacronym{few}{$\tt few$}{}
\newacronym{psd}{PSD}{}
\newacronym{snr}{SNR}{}
\newacronym{fm}{FM}{}
\newacronym{sbi}{SBI}{}
\newacronym{nf}{NF}{}

\usepackage{txfonts}
\begin{document} 

   \title{
  Simulation-Based Inference on EMRI Frequency Tracks: Harmonic Identification and Parameter Constraints from Search Outputs
   }
   \titlerunning{Simulation-Based Inference on EMRI Frequency Tracks}

   %\subtitle{I. Overviewing the $\kappa$-mechanism}

   \author{ F. Fantoccoli\inst{1} \and L. Speri\inst{2,3} \and B. Depoorter\inst{4, 5, 6} \and J. R. Gair\inst{1} } \institute{ Max Planck Institute for Gravitational Physics (Albert Einstein Institute), Am Mühlenberg 1, D-14476 Potsdam, Germany \email{federico.fantoccoli@aei.mpg.de} \and European Space Agency (ESA), European Space Research and Technology Centre (ESTEC), Keplerlaan 1, 2201 AZ Noordwijk, The Netherlands \and Leiden Observatory, Leiden University, P.O. Box 9513, 2300 RA Leiden, The Netherlands \and Department of Physics and Astronomy , KU Leuven, Celestijnenlaan 200D, B-3001 Leuven, Belgium \and Leuven Gravity Institute, KU Leuven, Celestijnenlaan 200D box 2415, 3001 Leuven, Belgium \and  Royal Observatory of Belgium, Avenue Circulaire, 3, 1180 Uccle, Belgium
   }
   \authorrunning{F. Fantoccoli et al.}

   \date{Received ---, accepted ---}

% \abstract{}{}{}{}{} 
% 5 {} token are mandatory
 
  \abstract
   {Extreme Mass Ratio Inspirals (EMRIs) are key targets for the Laser Interferometer Space Antenna, but their long, multi-harmonic waveforms make their search and parameter estimation computationally expensive and challenging with standard methods. 
   A promising alternative is to work directly with the output of a time–frequency search pipeline where one or more recovered frequency tracks are identified.}
   {We investigate whether frequency-evolution tracks recovered by a semi-coherent search pipeline can, on their own, be used to (i) identify which harmonic(s) a recovered track corresponds to; and (ii) constrain the intrinsic EMRI parameters, including an assessment of how these constraints degrade when the tracks are partially masked near plunge.}
   {We generate large banks of azimuthal and radial fundamental-frequency tracks and train Simulation-Based Inference (SBI) networks on single-frequency and combined inputs. 
   For harmonic identification, we combine pairs of candidate tracks under different harmonic-label assumptions, use the resulting posterior medians and network acceptance rates as summary statistics, and classify the correct labelling. 
   For parameter identification, we apply a sequential (truncated-prior) SBI scheme to progressively shrink the parameter volume, and further refine the results by cross-checking posterior samples against a time–frequency detection statistic.}
   {The harmonic-identification pipeline recovers the correct harmonic pair among the top-2 ranked candidates with probability $>0.99$ across masking levels, and correctly identifies it outright in 98\% of cases when restricted to physically plausible candidates. 
   The sequential SBI parameter-identification pipeline shrinks the intrinsic-parameter prior volume by up to $10^{13}$
   within a few hours of GPU time. 
   Combining the SBI posterior with a detection-statistic-based refinement step further reduces the constrained volume by up to four additional orders of magnitude, approaching the precision achievable in the least-masked cases even for heavily masked tracks.}
   {Frequency-track-based SBI provides a fast, waveform-free method to identify EMRI harmonics and to obtain tight, reliable prior volumes to enable targeted follow-up parameter estimation, substantially reducing the burden on subsequent MCMC-based analyses and offering a practical strategy for the multimodal, high-precision regime characteristic of EMRI parameter inference.
   }

   \maketitle
%-------------------------------------------------------------------
\section{Introduction}\label{sec:Intro}
The detection and characterisation of Extreme Mass Ratio Inspirals (\gls{emris}) represents one of the most interesting challenges in the context of Laser Interferometer Space Antenna (LISA) \citep{LISA_red_book} data analysis. \gls{emris} are binary systems in which a compact object (CO), most likely a black hole (\gls{bh}) with mass $m_2 \in [10, 100] \ M_{\odot}$, orbits a massive black hole (\gls{mbh}) with typical mass $m_1 \in [10^5, 10^7] \ M_{\odot}$, corresponding to mass ratios $q$ in the range $10^{-3}$--$10^{-6}$.
The gravitational wave (\gls{gw}) emission from \gls{emris} spans the frequency range $10^{-4}$--$10^{-1} \ \mathrm{Hz}$, making such systems a primary target for the LISA mission \citep{Babak_2017_EMRIscience,EMRI_science_uniquepot,Speri:2026ade}. Owing to their slow orbital evolution, a typical \gls{emri} remains in the LISA band for several years, during which it undergoes many orbital cycles, allowing extremely precise measurements of its physical properties.

These systems pose a number of data analysis challenges. 
First, the length and complexity of the \gls{gw} signal make waveform generation and likelihood evaluation expensive ($\sim 100 \ \mathrm{ms}$ for a 4 year long, $q = 10^{-5}$ signal \citep{Chapman_Bird_2025_FastFramDragging}), a cost that could become prohibitive when repeated over the many iterations required to search for and then infer the EMRI parameters.
% by a stochastic sampler. 
Second, because of the precision with which the parameters can be measured, the ratio between the posterior and prior volumes is $10^{-5}$--$10^{-6}$ per parameter, which makes fully-coherent matched filtering searches unfeasible~(\cite{Gair:2004si}).
Moreover, the complex trajectories followed by the CO in the gravitational field of the more massive body imply that the \gls{gw} signal is a superposition of many \textit{harmonics}, constructed from combinations of the orbital frequencies along the radial ($\Omega_{r}$), azimuthal ($\Omega_{\phi}$) and polar ($\Omega_{\theta}$) directions. 
The presence of multiple harmonics favours the emergence of non-local degeneracies \citep{Gair:2008zc,Babak:2009ua,Chua_2022} in parameter space, which make MCMC sampling extremely inefficient and potentially biased: at these degeneracies the posterior surface develops \textit{secondary} peaks, where the MCMC walkers tend to linger. 
It is worth stressing that the parameter space volume enclosed by the secondary modes is bigger than the one of the primary mode; however, the posterior volume covered by the secondary modes is significantly smaller than that of the main (true) mode. 
There are other related characteristics of the EMRI signal worth mentioning in a data analysis prospective. Firstly, the \gls{snr} accumulates slowly along the trajectory (much as for a continuous \gls{gw} signal), rather than in a burst close to the merger of the two compact objects, so that even when the signal is present in the data its instantaneous \gls{snr} is small. Secondly, the \gls{snr} is spread across multiple harmonics, depending on the intrinsic parameters of the system (mostly $e$ and $i$), so that each individual harmonic can carry only a small fraction of the total \gls{snr}. Both of these features have a great impact on the choice of the search strategy of such signals in a data stream, leaving plenty of room for improving data analysis strategies, and motivate exploring parameter estimation techniques beyond MCMC sampling and its variants. 

A variety of complementary approaches have been developed to address different aspects of the problem.
Within the context of MCMC methods, employing tailored annealing schedules, detection statistic modifications, and problem-specific proposal distributions can mitigate the impact of secondary maxima and improve exploration of the posterior distribution \citep{Speri_2026, Strub:2025dfs, Babak:2009ad, Cornish:2008ra}.
The use of veto likelihoods has also been proposed as a mechanism to suppress spurious local maxima and improve convergence towards the true signal parameters~\citep{Chua:2022ssg}.
Machine-learning techniques have emerged as promising tools for accelerating both detection and inference.
These methods have been applied to reduce waveform-generation costs~\citep{Zhao:2023ncy,Zhang:2022xuq,Yun:2023oqk}, improve sampling efficiency~\citep{Liang:2025vuf}, and perform rapid parameter estimation.
In particular, sequential simulation-based inference (SBI) techniques, such as Truncated Marginal Neural Ratio Estimation, have shown potential for efficiently constraining the high-dimensional EMRI parameter space starting from broad prior distributions~\citep{Cole:2025sqo}.
A distinct class of methods operates in the time--frequency domain. 
These approaches identify characteristic EMRI tracks in spectrograms without requiring the generation of waveform templates, thereby substantially reducing computational costs~\citep{Gair:2007ym,Gair:2006nj,Wen:2006vi}.
Related efforts have also employed phenomenological waveform models to construct template families capable of covering broad regions of parameter space.
Such approaches enable model-independent signal identification prior to parameter refinement with a physical waveform model~\citep{Wang:2012whc, Ye:2023xka,Speri_2026}.
Rather than developing a new search algorithm, we investigate how much astrophysical information can be extracted directly from the frequency tracks identified by a previous pipeline.
This has been investigated for a few systems in Refs.~\cite{Wang:2012whc, Ye:2023xka,Speri_2026}, but never systematically.
For this reason, we aim to understand whether the frequency evolution alone can provide useful constraints on the source parameters and thereby inform subsequent, computationally expensive waveform-based analyses.

Our main goal is to use the output of an \gls{emri} search  providing one or more frequency tracks to address the following questions:
\begin{enumerate} 
    \item \label{itm:harmonic-id} Given a frequency track obtained from a search pipeline, can we identify from the track alone which harmonic it corresponds to? 
    \item \label{itm:parameter-constraints} Assuming the harmonic is known and the track is recovered perfectly, how well can we constrain the parameters from the frequency evolution alone? 
    \item \label{itm:track-precision} If the search pipeline constrains a track only up to some finite precision, how do the constraints obtained above degrade? 
\end{enumerate}

This work is organized as follows. In Sec.~\ref{sec:methods} we introduce the strategy adopted to generate the noisy dataset, we also describe the general neural network infrastructures we use for this work. Then, we provide methods and results for each of the aforementioned questions.
In Sec.~\ref{sec:mode_selection} we discuss the procedure (Sec. \ref{sec:mode_selection_methods}) and the results (Sec. \ref{sec:mode_selection_results}) of the harmonic identification process (question \ref{itm:harmonic-id}). 
The same organization is proposed in Sec. \ref{sec:PE} where we describe our strategy to shrink the parameter space volume (questions \ref{itm:parameter-constraints} and \ref{itm:track-precision}); again divided in methods (Sec. \ref{sec:PE_methods}) and results (Sec. \ref{sec:PE_results} and \ref{sec:PE_improvements}). 
In the last section (Sec. \ref{sec:DnC}) we present our answers to the questions and conclude with the implications and future outlook.

%--------------------------------------------------------------------
\section{Methods}\label{sec:methods}

\subsection{Extreme mass ratio inspiral frequency evolution}\label{sec:EMRI_tracks}
We split the parameters that characterize an \gls{emri} \gls{gw} signal into two groups, \textit{intrinsic} and \textit{extrinsic}. The detector-frame primary and secondary masses, $m_1$ and $m_2$, the dimensionless spin parameter of the primary, $a$, the orbital eccentricity at plunge, $e_f$, the orbital inclination,, $i$, the initial azimuthal, radial and polar phases, ($\Phi_{\phi_0}, \Phi_{r_0}, \Phi_{\theta_0}$), and the time to plunge, $T_\mathrm{pl}$ from the start of observation are labelled as intrinsic, whereas the luminosity distance, $d_L$, together with the angular positions and orientations, $(\theta_k, \phi_k, \theta_s, \phi_s)$, are referred to as \textit{extrinsic} (see \cite{Chapman_Bird_2025_FastFramDragging} for details on the parameters). In this work we are mainly interested in constraining the intrinsic parameters. Furthermore, we restrict ourselves to equatorial, prograde \gls{emris}, in which the orbit of the secondary lies in the equatorial plane throughout the evolution, with its orbital angular momentum aligned with the spin of the primary.

As anticipated in Sec.~\ref{sec:Intro}, we work with single harmonic frequency evolution tracks, focusing on the frequency evolution alone and neglecting the information carried by the amplitude of the signal in its time--frequency representation. The intrinsic parameters of an \gls{emri} shape its frequency evolution changing its frequency span (primary mass dependence), its derivative (mostly determined by primary and secondary mass) and the growth in the lead up to the plunge phase (shaped by the dimensionless spin parameter and eccentricity). In this work we will focus on equatorial EMRIs. There are several reasons for this. Firstly, we expect eccentric and equatorial EMRIs to have fewer harmonics and thus be more amenable to the search and parameter approach proposed here. This therefore constitutes a good starting problem to gain insight before transferring the ideas to generic EMRIs. In addition, the EMRIs that will be included in the next LISA data challenge will be eccentric equatorial, allowing us to test our strategy on realistic data in that context. Finally, but also very importantly, the eccentric and equatorial waveforms available from the $\tt FastEMRIWaveform$ ($\tt few$) python package \citep{Katz_2021, 2023arXiv230712585S, chapman_bird_2025_15630565} are based on accurate numerically computed energy, angular momentum and carter constant fluxes, and accurate waveforms. When the EMRI orbit is non equatorial the current version of $\tt few$ relies on the Analytic Kludge approximation \citep{Chua_2017}, which is less accurate. A generic equatorial \gls{emri} frequency track $f_{m,n}$ can be decomposed as a function of the fundamental frequencies in Kerr spacetime as follow:
\begin{equation}
    f_{m,n}(t) = \frac{m \Omega_{\phi}(t) + n\Omega_r(t)}{2 \pi}, \label{eq:track_evol}
\end{equation}
where the pair $(m, n)$ identifies the harmonic and $\Omega_{\phi}(t)$, $\Omega_{r}(t)$ are the fundamental azimuthal and radial frequencies in Kerr spacetime \citep{Fujita_2009,schmidtCelestialMechanicsKerr2002}. For any \gls{emri} signal, both the number of relevant harmonics and the distribution of power among them depend strongly on the intrinsic parameters of the system, and in particular on the eccentricity \citep{Chapman_Bird_2025_FastFramDragging}. Throughout this work we compute the frequency evolution tracks with the $few$ python package \citep{Katz_2021, 2023arXiv230712585S, chapman_bird_2025_15630565}.

\subsection{Simulation-based inference}\label{sec:SBI}
We introduce our notation starting from Bayes' theorem,
\begin{equation}
    p(\theta|d) = \frac{p(d|\theta) p(\theta)}{p(d)}, \label{eq:Bayes_theorem}
\end{equation}
where $\theta$ is the $N$-dimensional array of parameters and $d$ the dataset, $p(d|\theta)$ is the likelihood function, $p(\theta)$ the prior on the parameters, $p(d)$ the evidence and $p(\theta|d)$ the posterior distribution. The machine learning strategy at the centre of this work is Simulation-Based Inference (\gls{sbi}) \citep{Cranmer_2020, deistler2025simulationbasedinferencepracticalguide}, originally proposed to tackle \textit{likelihood-free} Bayesian inference problems for which an analytical definition of the likelihood function $p(d|\theta)$ is not available. The likelihood evaluation is replaced by a set of simulations of synthetic data, each corresponding to a set of parameters $\theta$. During training, the \gls{sbi} network learns the posterior distribution by approximating it with a Normalizing Flow (\gls{nf}) \citep{papamakarios2018fastepsilonfreeinferencesimulation, papamakarios2021normalizingflowsprobabilisticmodeling}, and the loss function adopted is the negative log-likelihood of the data under the model. This choice can be shown to be equivalent to minimizing the \textit{Kullback--Leibler} (KL) divergence between the true underlying posterior $p(\theta|d)$ and the \gls{nf} approximation $q_\phi(\theta \mid d)$, where $\phi$ denotes the parameters of the network. Defining the loss function $\mathcal{L}$ as the expectation value of the KL divergence with respect to the data, we obtain
\begin{equation}\begin{split}
\mathcal{L} \equiv \mathbb{E}_{p(d)} D_{\mathrm{KL}}(p \mid\mid q_\phi)
=& \int \mathrm{d}d\, p(d) \int \mathrm{d}\theta\, p(\theta \mid d)\,
\log \frac{p(\theta \mid d)}{q_\phi(\theta \mid d)}\\
= & \int \mathrm{d}\theta\, p(\theta) \int \mathrm{d}d\, p(d \mid \theta)\,
\log \frac{p(\theta \mid d)}{q_\phi(\theta \mid d)}\\
\approx& -\frac{1}{N} \sum_{\substack{\theta^{(i)} \sim p(\theta) \\ d^{(i)} \sim p(d \mid \theta^{(i)})}}
\log q_\phi(\theta^{(i)} \mid d^{(i)}) + \text{const.}
\label{eq:SBI_demo}
\end{split}\end{equation}
In practice, we draw samples from the prior distribution $p(\theta)$ of the physical parameters $\theta$ and use them to produce the simulated datasets $d$ (i.e., the likelihood draws), then we use such draws to minimize the KL divergence in Eq. (\ref{eq:SBI_demo}) and train the NF approximation of the posterior $q_{\phi}(\theta|d)$. We provide some references to applications of SBI in a gravitational-wave context: \cite{Green:2020hst,Green:2020dnx,Dax_2021, Spadaro_2026, cole2026sequentialsimulationbasedinferenceextreme, mao2025robustscalablesimulationbasedinference, lemke2026detectinggravitationalwaveanisotropiessimulationbased}. In all these works a variety of SBI strategies have been adopted for search and parameter estimation purposes. As far as EMRIs are concerned, it has been proven in \cite{cole2026sequentialsimulationbasedinferenceextreme} that applying SBI to EMRI waveforms is challenging due to their high complexity and long duration. Both these factors contribute to increase significantly the computational cost of the simulation and training process, requiring also a large amount of memory. The search for an alternative approach with respect to the computationally expensive waveforms is one of the motivations behind this work.

\subsection{Data generation process}\label{sec:dataset}
In this section we describe the steps we follow to reproduce the output of a time-frequency search pipeline. We describe the data generating process to draw parameters from the prior $\theta \sim p(\theta)$ in order to then obtain a frequency track $f(t;\theta)$, and then add noise $\delta f$ to produce the data $d$. We start by drawing the intrinsic parameters of the system, $\theta_\mathrm{int} = (m_1, m_2, a, e_f, T_\mathrm{pl})$, from the prior distributions listed in Tab. \ref{tab:priors}.
The prior boundaries for each parameter are comparable to the expected ranges from astrophysical EMRI populations \citep{Babak_2017_EMRIscience}. In this work we focus on prograde orbits, however the generalization to include retrograde systems is simple and requires an extension of the spin parameter prior between $a \in [-0.99, 0.99]$. The size of the bank depends on the application, and for this work we use $5 \times 10^{5}$ parameter sets. We use the $\tt fewtrax$ python package (in preparation \cite{Fewtrax}) to compute the fundamental frequencies $\Omega_{\phi}(t)$ and $\Omega_r(t)$ associated with each parameter set $\theta$. This generation step produces $5 \times 10^5$ frequency evolutions in $O(1)$ minute on a laptop CPU. We generate two year long frequency tracks with a cadence of $\sim 0.6$ days, so that each fundamental frequency evolution consists of 1262 points spanning the interval $[0, 2] \ \mathrm{yr}$ and mimicking the resolution of a time--frequency grid. Shorter signals ($T_\mathrm{pl} < 2 \ \mathrm{yr}$) are simulated by \textit{rolling} the frequency track backwards, so that the plunge (the maximum frequency point) occurs at $T_\mathrm{pl}$, and padding the evolution with zeros for $t > T_\mathrm{pl}$.

\begin{table}
    \centering
    \begin{tabular}{l l l}
    \toprule
    \textbf{Parameter} & \textbf{Prior} & \textbf{Units} \\
    \midrule
    $\mathrm{log}_{10}m_1$        & $\mathcal{U}(5, 7)$ & $M_{\odot}$ \\
    $m_2$        & $\mathcal{U}(10, 100)$      & $M_{\odot}$ \\
    $a$          & $\mathcal{U}(0, 0.99)$      & --- \\
    $e_f$        & $\mathcal{U}(0.005, 0.3)$   & --- \\
    $T_\mathrm{pl}$   & $\mathcal{U}(0.1, 2.0)$ & yr \\
    $f_\mathrm{mask}$ & $\mathcal{U}(0, 0.4)$   & --- \\
    \bottomrule
    \end{tabular}
    \caption{Prior distributions adopted for the generation of the training
    set. All priors are uniform within the quoted boundaries. The masses
    $m_1$ and $m_2$ refer to the primary and the secondary respectively,
    $a$ is the dimensionless spin of the primary, $e_f$ the eccentricity at
    plunge, $T_\mathrm{pl}$ the time to plunge measured from the beginning of
    the two year observation window, and $f_\mathrm{mask}$ the fraction of the
    track masked by the search pipeline close to plunge. The spin is
    restricted to positive values since we consider equatorial, prograde
    orbits only.}
    \label{tab:priors}
\end{table}

The output of a search pipeline is a frequency evolution that approximates the true track contained in the dataset. The quality of the track recovery depends on the performance of the pipeline itself.
A careful characterization of the pipeline on a realistic dataset therefore provides an estimate of the typical spread of the relative difference as a function of time:
\begin{equation}
    \delta f(t) = \frac{(f_{r}(t) - f_{T}(t))}{f_{T}(t)},
\end{equation}
where $f_r(t)$ is the recovered frequency track, while $f_{T}(t)$ is the true injection. 
In this work we refer to the semi-coherent search strategy presented in \cite{Speri_2026}, for which $\delta f(t)$ is of the order of $10^{-3}$, as also found in the work of~\cite{Wang:2012whc}. There are two main sources of uncertainty in $\delta f(t)$. %and is of order $10^{-3}$ also similar to \cite{Wang:2012whc}.
The first is the recovery uncertainty in the presence of noise: the semi-coherent approach matches each chunk of the time--frequency representation of the waveform independently discarding the information encoded in the coherent phase evolution across chunks; 
if the signal is instantaneously weak in every chunk the recovered frequency evolution may be biased towards randomly loud noise features. 
The second is related to an assumption built into the pipeline \citep{Speri_2026}, namely that the frequency evolution is linear within each chunk: close to plunge, if the amount of frequency evolution within a chunk exceeds a fixed threshold, this assumption breaks down and the pipeline output beyond that time is discarded (padded to zero). 
In previous works \citep{Speri_2026, tenorio2025scalabledataanalysisframeworklongduration}, the threshold was chosen to be: $ \ddot{f}_{\alpha} T^3_{
SFT}/6 > 1 $, where $T_{SFT}$ is the time length of a short fourier transform chunk and $f_{\alpha} $ the value of the frequency in chunk $\alpha$. 
We refer to this behaviour as \textit{masking} of the plunge, and the resulting masked fraction depends entirely on the \gls{emri} parameters. In particular, the last visible time sample of the frequency evolution ($T_{vis}$), the plunge time ($T_{\rm pl}$) and the masking fraction ($f_{mask}$) are related by the following expressions:
\begin{equation}
    T_\mathrm{vis} = ( 1 - f_\mathrm{mask}) \ T_\mathrm{pl}, \qquad f_\mathrm{mask} =  1 - \frac{T_\mathrm{vis}}{T_\mathrm{pl}}. \label{eq:fmask_analytic}
\end{equation}
These relations are discussed in greater detail in Sec. \ref{sec:NN_training}. An example of a frequency track after the rolling and masking is presented in the top panel of Fig. \ref{fig:residuals_and_Trackmask}. The figure shows the true fundamental frequency evolution (dotted line) from $\tt few$. The frequency evolution undergoes the process of rolling (blue line), where the $T_{PL}$ is adjusted from two years to the chosen value (1.9 years in this case), then, to obtain the masked frequency evolution (orange line), the mask is applied and the final $f_{mask}$ fraction of the frequency evolution is padded to zero (shaded area). To estimate the distribution of $\delta f(t)$ we collect the residuals from $300$ search pipeline runs on different signal injections at $\mathrm{S/N} = 30$ from the results in \cite{Speri_2026}. We compute the residuals up to the masking time (which in some cases coincides with the plunge time); because the amount of masking in each track depends on the physical parameters that describe the track, the length of the residuals are different across the 300 cases. Therefore, we normalize each residual onto a common track fraction grid spanning the interval $[0, 1]$, where 0 corresponds to the beginning of the signal and 1 to plunge. The resulting residual band is shown in the bottom panel of Fig. \ref{fig:residuals_and_Trackmask}. We use this band to map realistic search pipeline uncertainties onto a different set of generated tracks as:
\begin{equation}
    f_\mathrm{noisy}(t) = f_\mathrm{clean}(t)\left[1 + \delta f(t)\right], \label{eq:noisy_tracks}
\end{equation}
where $f_\mathrm{clean}$ is the track generated with $\tt fewtrax$ masked and rolled as explained above, meanwhile $f_\mathrm{noisy}$ is our model of the realistic output of a noisy search. The $\delta f(t)$ band is obtained by overlaying the 300 realizations $\delta f_i(t)$, which can be represented through a \textit{Singular Value Decomposition} (SVD) with two components. The SVD basis encodes the time evolution of the $\delta f_i(t)$ and can therefore be used to generate new, independent realizations. We find that the optimal way to replicate the residuals $\delta f$ from \cite{Speri_2026} and be able to extract new ones is to produce a `jitter' around the true ones with the SVD basis. We generate a new residual as follow:
\begin{equation}
    \delta f_{\rm new} = \delta f_{\rm search} + \varepsilon \ \sum_{i=0}^2 a_i \psi_i.
\end{equation}
Where $\delta f_{search}$ are the stored residuals from the search, $\varepsilon $ is a rescaling factor that we set to $0.2$, $a_i$ are the SVD coefficients drawn from a Standard Normal distribution and $\psi_i$ are the SVD basis functions. The results of the noise generation are illustrated in Fig. \ref{fig:residuals_and_Trackmask}. As can be seen from the figure, the residuals from the simulations matches the distribution obtained from the search. Both the distributions are strongly peaked around small $\delta f$ values, with median $\delta_f \simeq 10^{-3}$ (see Fig. 4 in \cite{Speri_2026}), however the distribution exhibits long tails up to $\delta f \sim 1.5$, especially in the early part of the recovered tracks.
\begin{figure}[h!]
    \includegraphics[width=\linewidth]{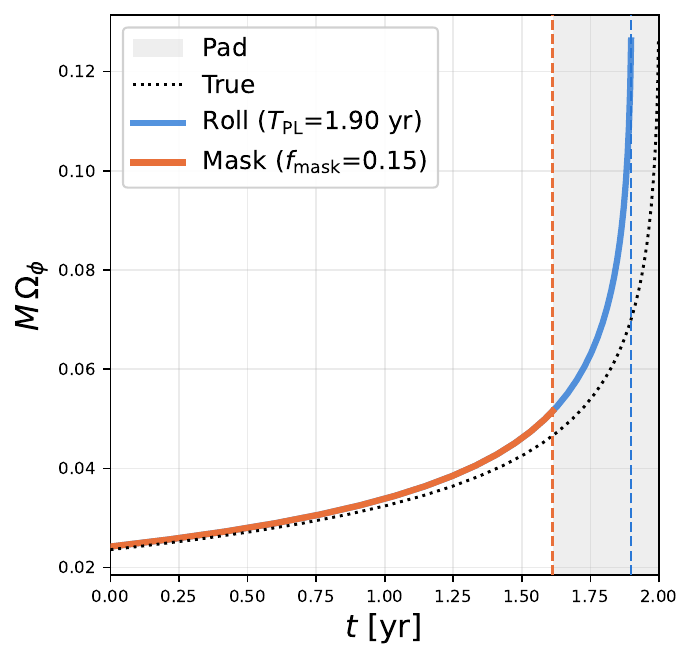}
    \caption{The 2-years long dimensionless fundamental frequency evolution track $M \Omega_{\phi}$, in dotted black line. $M$ is the primary mass of the EMRI, in solar masses. The blue line is the same track after the rolling procedure, and the orange track is the track after the masking has been applied, as described in Sec. \ref{sec:dataset}}
    \label{fig:trackmask_only}
\end{figure}

\begin{figure}[h!]
    \includegraphics[width=\linewidth]{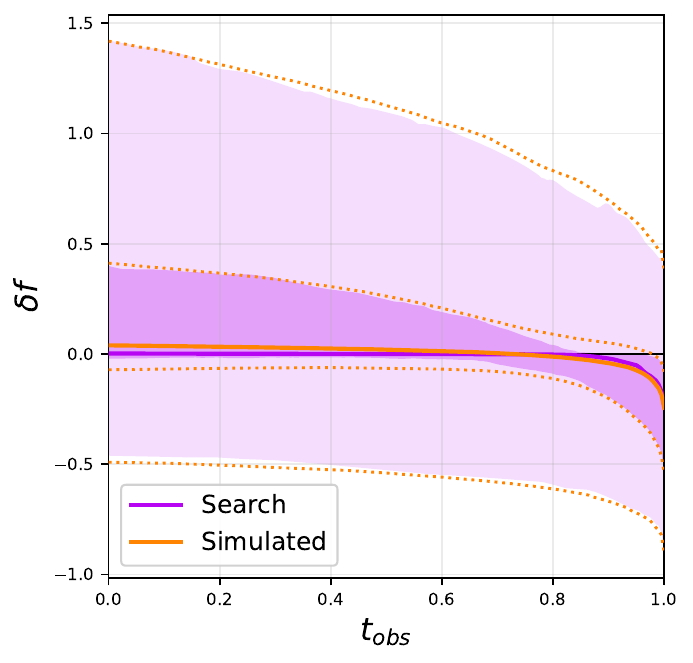}
    \caption{The $\delta f$ median (thick purple line) and the quantiles $25\%-75\%$ and $5\% - 95\%$ from the EMRI search pipeline \citep{Speri_2026} as shaded purple areas. The x-axis $t_{obs}$ is the observed fraction of the track from the beginning of the evolution (0) to the last visible point (1). The continuos orange line is the median of our simulated residuals, while the dotted lines are the $25\%-75\%$ and $5\% - 95\%$.
    }
    \label{fig:residuals_and_Trackmask}
\end{figure}

With the ingredients above we build a training set of frequency evolutions by randomly sampling $5 \times 10^5$ parameter sets $\theta_\mathrm{int}$ from uniform distributions over the ranges given in Tab.\ref{tab:priors}. We then roll each two years track to the sampled time to plunge, apply the mask by drawing the masked fraction uniformly within the interval reported in Tab. \ref{tab:priors}, and produce the noisy version of the fundamental frequency tracks through $\Omega_{i, \mathrm{noisy}}(t) = \Omega_{i, \mathrm{clean}}(t)\left[1 + \delta f(t)\right]$.

We stress that the relative uncertainty band of the search pipeline is computed on single harmonic frequency evolutions, which are linear combinations of $\Omega_{\phi}$ and $\Omega_r$ as in Eq.~\eqref{eq:track_evol}. Given two harmonic tracks and the corresponding pairs of harmonic numbers $(m_a, n_a)$ and $(m_b, n_b)$, however, the fundamental frequencies can be recovered as
\begin{equation}\begin{split}
    \Omega_{\phi}(t) &= 2\pi\,\frac{n_b f_{m_a, n_a}(t) - n_a f_{m_b, n_b}(t)}{m_a n_b - n_a m_b}, \\[5pt]
    \Omega_{r}(t)    &= 2\pi\,\frac{m_a f_{m_b, n_b}(t) - m_b f_{m_a, n_a}(t)}{m_a n_b - m_b n_a}, \label{eq:fund_freq_equations}
\end{split}\end{equation}
so that, assuming the same relative uncertainty for $f_{m_a, n_a}$ and $f_{m_b, n_b}$, it propagates directly to the fundamental frequencies: $\Omega_{i, \mathrm{noisy}}(t) = \Omega_{i, \mathrm{clean}}(t)\left[1 + \delta f(t)\right]$. 
We thus assume that both the fundamental frequency evolutions are affected by the same noise realization. This assumption is relevant in the case two tracks have been obtained in the early stage of the search, and it can be relaxed. In the context of the Mojito LISA data challenge analysis, we relaxed this assumption obtaining consistent results. This assumption could be relaxed given a better understanding of the pipeline uncertainties on different harmonics (both $m_a, n_a$, and $m_b, n_b$), which we do not investigate here. Moreover, we emphasise that if we extend the simulated quantile band to $\sim 98 \%$ quantile in Fig. \ref{fig:residuals_and_Trackmask}, we find a $\delta f \leq -1$ for $t_{obs} \simeq 1$,  leading to negative fundamental frequency samples from $\Omega_{i, \mathrm{noisy}}(t) = \Omega_{i, \mathrm{clean}}(t)\left[1 + \delta f(t)\right]$. Given our assumptions about the noise linearity between $f$ and $\Omega$, such noise realizations are not possible outcomes of a search pipeline where the recovered frequency $f_{\rm noisy}$ is positive at every time. It is possible to clip the noise band $\delta f $ to a maximum value $\mathrm{min}(\delta f) > -1$, however we find that applying such clipping makes no significative difference to our results.

\subsection{Neural Networks and Training}\label{sec:NN_training}

We use the bank of noisy fundamental frequency realizations to train a set $(d = \Omega_{i, \rm noisy}(t))$ of three \gls{sbi} networks. Two of them are trained on a single fundamental frequency evolution each: we label the one trained on the azimuthal frequency, $\Omega_{\phi}$, the \textit{azimuthal} network and the one trained on the radial frequency, $\Omega_{r}$, the \textit{radial} network, and refer to both as the \textit{single frequency} networks. The third network is trained on both frequencies simultaneously, and we label it the \textit{composite} network. The training set of the single frequency networks consists of the pairs $(d, \theta) = (\Omega_{i j, \mathrm{noisy}}(t), \theta_{j})$, where $i \in \{\phi, r\}$ and $j$ labels the simulation, each track being 1262 points long. The training set of the composite network is the concatenation of the azimuthal and radial frequency evolutions, i.e., an array of $1262 \times 2 = 2524$ elements.

Each \gls{sbi} network is produced and trained with the $\tt sbi$ python package \citep{tejero-cantero2020sbi}. Every network trained is preceded by an embedding Multi-Layer Perceptron (MLP) that compresses the time series to a lower dimensionality; here we adopt a six hidden-layer MLP that reduces the tracks from 1262 to 128 features. We find the performance not to depend significantly on the shape of the MLP, and defer a more systematic investigation of alternative representations of the frequency tracks to future work. The normalizing flow architecture adopted is a \textit{Neural Spline Flow} \citep{durkan2019neuralsplineflows} with 8 transforms and 64 hidden features. We train on batches of $\sim 8 \times 10^3$ samples with a learning rate of $5 \times 10^{-4}$ and a cosine annealing scheduler that drops the learning rate to $1 \times 10^{-6}$ in 1600 epochs. We reserve $90\%$ of the dataset for training and $10\%$ for  validation, and stop the training once the loss function has not improved for 100 epochs or when the maximum number of epochs (1600) has been reached. We underline that we tested two different training options: in the first case scenario, we re-sample the residuals, $\delta f$, to apply to the clean, rolled and masked, fundamental frequencies, $\Omega_{i, \rm clean}(t)$, for both the training and validation set at every epoch; in the second case we also resample the masking fraction accordingly. In both cases we are helping the neural network to marginalize over different noise realizations and extract the physical information contained in the track. We find that the combined resampling of noise and masking performs slightly better than only noise resampling for the neural networks described above, providing larger p-values associated with the P-P plots (see Appendix \ref{sec:Appendix_A}). A typical training run takes a few hours on a GPU. 
After the training, given a noisy set of fundamental frequencies, the networks predict the posterior distribution of the physical parameters, $\theta = (m_1, m_2, a, e, T_\mathrm{pl})$. Although the networks learn the distribution of possible masking fractions, this quantity is treated as a nuisance parameter and is not inferred. We note that the masking fraction is degenerate with $T_\mathrm{pl}$ because of the way the two are encoded in the training set, being related analytically by Eq. \eqref{eq:fmask_analytic}. The networks can partially break this degeneracy by exploiting the last non-zero value of the frequency evolution, since a larger masking fraction generally implies a lower final observed frequency. The degeneracy could be broken more easily by adding the frequency derivative at the last non-zero point to the training input, as a high masking fraction makes it more likely that the last available point lies far from plunge, in the early inspiral; we do not include this extra information here. The impact of frequency masking is studied in more detail in Sec. \ref{sec:PE}.

\section{Harmonic labelling Problem}\label{sec:mode_selection}
\subsection{Methods}\label{sec:mode_selection_methods}
In this section we describe a possible strategy to address question \ref{itm:harmonic-id} of Sec.~\ref{sec:Intro} using the \gls{sbi} networks introduced above. The most general frequency track of an equatorial \gls{emri} is given by Eq.~\eqref{eq:track_evol}, so that each track is characterized by an unknown combination of fundamental angular frequencies ($\Omega_{\phi}, \Omega_r$) given by harmonic numbers $(m, n)$. Since a generic \gls{emri} signal contains several harmonics, we expect a search pipeline to recover the most powerful tracks, returning a set of frequency evolutions $\{f_{m_1, n_1}(t), f_{m_2, n_2}(t), \dots, f_{m_N, n_N}(t)\}$.

Any two such tracks can be combined to obtain the underlying fundamental frequency evolutions through Eq.~\eqref{eq:fund_freq_equations}. Given two observed tracks $f_{m_a,n_a}$ and $f_{m_b,n_b}$, we \textit{assume} a pair of harmonic labels $(m_a, n_a)$ and $(m_b, n_b)$ and plug them into Eq. \ref{eq:fund_freq_equations} accordingly, to obtain the corresponding fundamental frequencies. If the assumed labels are correct, the frequencies obtained are the true fundamental frequencies of the signal, associated with the true underlying physical parameters. Looping over the possible label assignments, we build a set of candidate fundamental frequencies $(\Omega_{\phi, ab}(t), \Omega_{r, ab}(t))$, where the indices $a, b$ label the harmonic pairs $\{ (m_a, n_a), (m_b, n_b)\}$ under consideration. At this stage, we check that the fundamental frequencies obtained for each harmonic candidate are positive at every time; negative fundamental frequencies are immediately discarded.

We then use the single frequency networks to infer the parameters associated with each candidate azimuthal and radial evolution. If the harmonic labelling is correct, the two independent networks should support the same region of parameter space; if it is wrong, they should instead constrain different regions. The composite network behaves even more sharply: when the labelling is wrong it may struggle to return a posterior within the prior support, resulting in an extremely low (nearly zero) acceptance rate when sampling the posterior through the forward network pass. Given an observation, the acceptance rate is defined as the fraction of samples produced by the SBI networks forward pass that falls within the prior range. This happens because the combination of fundamental frequencies passed as input is often, though not always, unphysical, in the sense that it cannot be generated by any single, coherent set of intrinsic parameters of the EMRI $\theta$. A more extensive discussion on this point is provided in Appendix \ref{sec:Appendix_B}.

\begin{figure*}
    \centering
    \includegraphics[width=\linewidth]{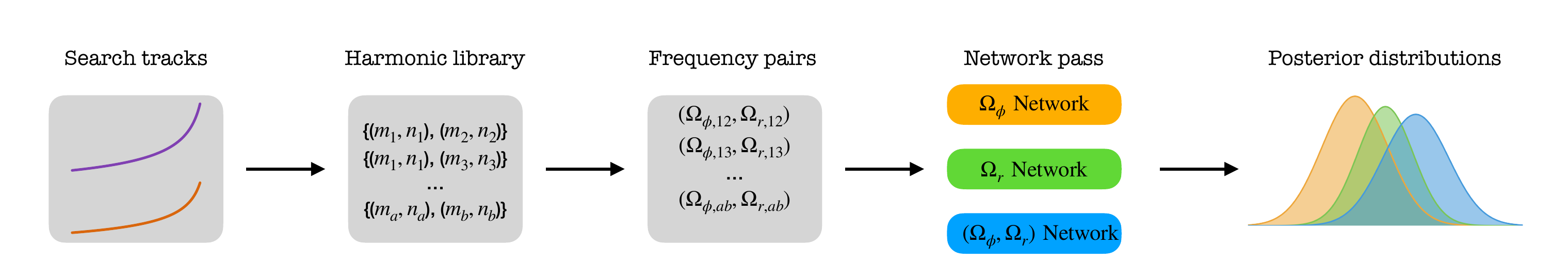}
    \caption{The schematic representation of the main steps of the mode selection process. We start by taking the tracks from the search pipeline, generating a harmonic library, looping over different pairs of harmonic sets $\{(m_a, n_a),(m_b, n_b)\}$. Then, we build fundamental frequency track pairs associated to the chosen harmonics ($\Omega_{\phi,ab}, \Omega_{r,ab}$) from Eq. \eqref{eq:fund_freq_equations}. We pass each of the fundamental frequencies in the corresponding neural network and we recover the posterior distributions. We collect medians and build the \textit{summary set}.}
    \label{fig:Diagram}
\end{figure*}

Given a pair of single harmonic frequency evolutions, we loop over the plausible harmonic label assignments and, for each of them, pass the reconstructed fundamental frequencies through both single frequency networks and the composite network, obtaining three sets of posterior samples in the physical parameters. For every assignment we then collect the medians of the three posteriors for each parameter, together with the acceptance rate of the composite network; we refer to these quantities as the \textit{summary set}. The acceptance rate plays an important role: if it falls below a given threshold, we stop sampling and classify the tested label assignment as \textit{inconsistent}. As explained above, the predictions of the three networks are expected to agree with one another only when the harmonic labels are the correct ones. A schematic representation of the process is provided in Fig.\ref{fig:Diagram}

In order to distinguish correct from incorrect label assignments, we use the summary set to train a classifier. We generate a mock dataset of recovered harmonic tracks as follows. We choose 10 masking levels, uniformly spaced in the interval $f_\mathrm{mask} \in [0.0, 0.37]$, and for each of them we generate 100 pairs of harmonic tracks $(f_{m_a,n_a}, f_{m_b,n_b})$, drawing the parameters from the priors of Tab.\ref{tab:priors}, except for the time to plunge, which is fixed to $T_\mathrm{pl} = 1.9$ yr. In all injections the true harmonics are $(2, 1)$ and $(2, 2)$, and each of the 100 injections is generated with a different noise realization. For each pair of tracks we test 25 label assignments built from the harmonics $\{(2,0),(2,1),(2,2),(3,0),(3,1),(3,2),(4,0),(4,1),(4,2)\}$, so that every injection yields 24 \textit{inconsistent} assignments, which return incorrect fundamental frequencies through Eq.~\eqref{eq:fund_freq_equations}, and one \textit{consistent} assignment, which recovers the true underlying evolutions. In total we obtain $10 \times 100 \times 25 = 25000$ candidate fundamental frequency pairs to be fed to the networks. Each forward pass takes $\mathcal{O}(1)$ second to produce a posterior, so the whole procedure takes $\mathcal{O}(1)$ hours; we ran it on GPU, but we expect no significant difference on CPU. The true label pair chosen is not important because the neural networks take as input the true fundamental frequencies, that are the same for every choice of the true label (see Eq.\ref{eq:fund_freq_equations}). However, we underline that given the same set of 25 possible combinations, the results can change (improve in our case) with a change of labelling. The reason is that given the new set of true $(f_{m_a,n_a}, f_{m_b,n_b})$, the same 25 labelling combinations can provide a negative $\Omega_{\phi}(t)$ or $\Omega_r(t)$, that is immediately discarded by the pipeline, reducing the effective number of inconsistent candidates and an easier classification task. More details are provided in Appendix \ref{sec:Appendix_B}.

The classifier adopted is the $\tt ExtraTreesClassifier$ \citep{Extreme_threes} of the $\tt sklearn$ python package \citep{scikit-learn}, an extension of the Random Forest classifier \citep{Rand_forest}. Several classifiers available in $\tt sklearn$ are easily applicable to this problem; we tested a few of them and retained the one giving the best accuracy on our dataset. Each training example consists of four features: the three posterior medians, one per network, and the acceptance rate of the composite network. When the acceptance rate is too low to produce samples, we set it to zero and assign default values to the medians, $\mathrm{med}(\log_{10} m_1, m_2, a, e_f, T_\mathrm{pl}) = (0, 0, -1, -1, 0)$, chosen to lie outside any physical range of interest (see Tab.\ref{tab:priors}).

We train the classifier with a \textit{K-fold} strategy, to classify all 25000 label assignments. We adopt $K = 5$ folds, each splitting the dataset into a training set ($80\%$) and a validation set (the remaining $20\%$), with non-overlapping validation sets so that every element of the dataset is classified exactly once by a network (the $K$ classifier) that has not seen it during training. The output of the classifier can be interpreted as the probability that a given label assignment is the correct description of the underlying frequency evolution. For each of the $N_\mathrm{tracks} = 10 \times 100 = 1000$ track pairs, we rank the 25 tested assignments by classification score and record the position of the true one, so that each injection is associated with a rank $R \in [1, 25]$. We then compute the probability that the true assignment, for a fixed masking fraction, falls within the top $N$ scores as:
\begin{equation}
    P(\mathrm{true} \in \mathrm{top}\text{-}N) = \frac{\sum_i^{N_\mathrm{tracks}(f_{mask})} X_i}{N_\mathrm{tracks}(f_{mask})}, \quad X_i =\begin{cases}\label{eq:P_consistent}
    1 & \mathrm{if} \ R \leq N,\\
    0 & \mathrm{if} \ R > N.
    \end{cases}
\end{equation}
Where $N_{tracks}(f_{mask})) = 100$ is the total number of tracks given a fixed masking fraction.

\subsection{Results}\label{sec:mode_selection_results}
\begin{figure}
    \includegraphics[width=\linewidth]{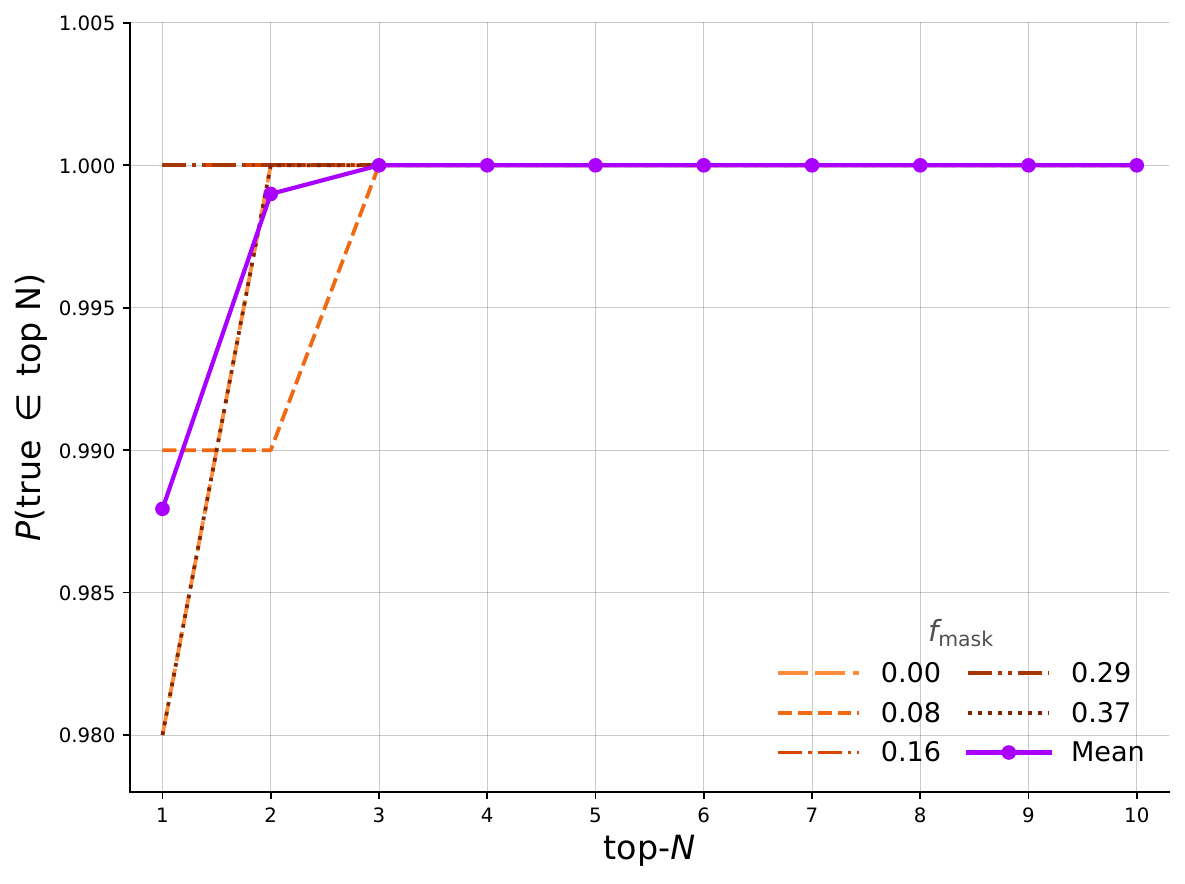}
    \caption{The x-axis shows the classifier rank to which we are including harmonics. On the y-axis the probability computed as illustrated in Eq. \eqref{eq:P_consistent}. The colours illustrate a different level of masking of the tracks, while the purple line is the average across all the masking fractions}
    \label{fig:Harmonic_topN}
\end{figure}

We display the probability computed through Eq. \ref{eq:P_consistent} in Fig. \ref{fig:Harmonic_topN}. We find that the curve rises steeply, providing a robust identification of the true candidate in most cases. As expected, the acceptance rate of the composite network is a powerful feature, as it discards many of the wrong assignments ($\sim 80 \ \%$) at an early stage (see also Appendix \ref{sec:Appendix_B}). Training the classifier on the medians alone, conditioned on the acceptance as described above, yields results very close to those of the full four-feature training. As can be seen from Fig. \ref{fig:Harmonic_topN}, the average performance (purple curve) across all the masking fractions selected is always above $\sim 0.98$, assuring a good overall performance across the masking fractions. The probability rises to $1$ at $N = 3$. The curves we illustrate in Fig. \ref{fig:Harmonic_topN} are the outcome of 100 different parameter sets and noise realizations for each mask. To assess the scatter of the curves due to randomized noise realizations and across parameter space we select three masking levels: $f_{mask} = \{0.0, 0.12, 0.37\} $, generate 2000 different tracks at each masking fraction chosen with parameters drawn from the astrophysical prior in Tab. \ref{tab:priors} and random noise realizations as described in Sec. \ref{sec:methods}. We compute the summary statistics, compute the ranks of each candidate and the probability at different $N$. We used the 2000 samples to provide 20 independent estimates of the probabilities in Fig. \ref{fig:Harmonic_topN} up to $N = 3$, finding that the standard deviations across 20 realizations is $\sim 0.01$ for $N = 1$ and $\lesssim 0.02$ for $N > 1$. We explore what is the highest source of confusion in our mode selection pipeline in Appendix \ref{sec:Appendix_B}.

We stress that these results are obtained by combining only two of the three fundamental frequencies $\Omega_{\phi}, \Omega_{r}, \Omega_{\theta}$. Including the polar frequency $\Omega_{\theta}$ would further enhance the discriminating power, pushing the top-1 probability closer to unity thanks to the larger number of available network combinations: a single \textit{polar} network, a \textit{fully combined} network using all three fundamental frequencies, and six \textit{two-frequency} networks. Although this may seem a large number of networks, we recall that the training cost is incurred only once, and amounts to a few hours on a single GPU for each of them. 

\section{Constraining the parameters}\label{sec:PE}
\subsection{Methods}\label{sec:PE_methods}
In the previous section we described a strategy for the mode selection problem. In this section we focus on the estimation of the intrinsic EMRI parameters through \gls{sbi}. As underlined in Sec. \ref{sec:Intro}, the goal is to use only the frequency evolution of an EMRI signal without dealing with slow and memory-expensive waveform generation. 

The neural networks trained in the previous section were computed adopting astrophysical prior ranges (see Tab. \ref{tab:priors}). 
%Although wide prior ranges reflects on the ability of the network to produce reliable posteriors across a wide parameter space, they (often) imply that the posterior obtained is broad
Wide prior ranges reflect the reality that the parameter space of possible astrophysical signals is large and the parameters of individual events unknown. However, with limited training data, SBI methods can produce overly broad posterior distributions when trained with broad priors, due to insufficient density of systems in the training set. 
The reason is the \textit{sampling density} of the training data across parameter space, defined as $D = N_{samp}/V_{prior}$, where $N_{samp}$ is the number of training samples and $V_{prior}$ is the prior parameter space volume. For the purpose of this work, we define $V_{prior} = \Pi_{\theta_i} |\theta_{R,i} - \theta_{R, i}|$, where $\theta_{L, i}$ and $\theta_{R, i}$ are the left-hand side and right-hand side of the prior edge for each parameter, indexed by $i$. 
The volume shrinking is extremely important in the context of EMRIs, where the posterior of intrinsic parameters is extremely sharp, with a posterior-to-prior one dimensional volume ratio that can be of the order of $10^{-4}-10^{-6}$, and the overall volume compression required for the whole posterior is $(10^{-4})^5 \sim10^{-20} $ with respect to the prior volume (the needle-in-a-haystack problem \citep{Gair:2004si,Strub:2025dfs,Speri_2026}. To achieve such a level of precision we make use of a sequential-\gls{sbi} approach (also applied in \cite{Speri_2026} and \cite{cole2026sequentialsimulationbasedinferenceextreme}), where starting from the prior (round-0 prior) we draw $N_{samp}$ samples from it, train an \gls{sbi} neural network to predict the posterior and use the information from the predicted posterior to select a smaller range of parameter space as the new prior. We repeat the process starting from the new round-1 prior in such a way that at every iteration we increase the density of the sampling space around the region of interest. In the absence of noise in the observation (and in training data) and provided a sufficiently dense training dataset, we find that the posterior from a sequential-\gls{sbi} can be shrank to obtain an arbitrarily accurate point estimate of the intrinsic parameters. However, in a realistic scenario the procedure can be repeated up to the point where the shrinking of the parameter space is stalled and no more information can be extracted from the data. The procedure requires some care: to obtain a proper, reliable, posterior distribution from a sequence of \gls{sbi} iterations, the prior chosen at each round should be a truncation of the original round-0 prior bounds (see \cite{deistler2022truncatedproposalsscalablehasslefree, TMNRE} for more details). A different prior in further rounds can be adopted, but the final posterior distributions must be corrected accordingly.  We choose the sequential SBI approach instead of a MCMC run in waveform space because of the multimodal structure of the posterior across parameter space: \cite{Chua_2022} showed that it is possible to explore the multimodality through a series of MCMC runs in the identified volume of parameter space. However each run could take several days to converge and provide densely sampled posterior distributions.

The starting point of the pipeline are the noisy frequency evolutions. Assuming we have identified the correct harmonic (with the procedure described in Sec. \ref{sec:mode_selection}), we generate the noisy fundamental frequency evolutions through Eq. \eqref{eq:fund_freq_equations}. We provide a round-0 posterior passing the noisy frequency evolution to the composite neural network trained on the large astrophysical prior. From the posterior outcome of this first pass we define the new prior as follows. Using the samples in the sampler space, we run a \textit{Principal Component Analysis} (PCA) to identify the main correlation directions of the EMRI parameter space, then we project the samples along such directions and build a box around each projected marginal posterior of every parameter that encloses the $1\% - 99\%$ quantiles of the distribution. The boundaries of the box, projected back to the physical space define the prior on each parameter for the following round. This prior choice step is particularly important allowing the subsequent rounds to focus on the most likely region of parameter space that describe the observed track. The round-0 EMRI posterior across parameter space is on scales $\sim 10 \%$ of the true parameter, and comprises strongly correlated ridges (for instance, see Fig. \ref{fig:SBI_shrinking_posterior}, or Fig. 6 in \cite{Speri_2026}) and we can use such ridges to guide the next rounds' prior choices. In this work we are not focussed on predicting the true posterior distribution of the intrinsic parameters, but rather to localize them to the smallest possible \textit{identification volume} within which the true parameters of the signal might lie.  Such information could then be used to initialize (or propose points for) a Markov Chain Monte Carlo sampler run to estimate the parameters with the required level of precision. For this reason, we are not correcting for a different prior choice through the rounds. Alongside the physical prior changes, we also change the prior on the $f_{mask}$ nuisance parameter, effectively reducing the $f_{mask}$ fractions allowed in the next rounds. The reason is that the $f_{mask}$ parameter is completely degenerate with $T_{PL}$, so the posterior learned by the neural network would be the posterior from the $T_{PL}$ transformed by Eq. \eqref{eq:fmask_analytic}. At round-0 we fix the new allowed uniform window to use as next rounds' prior as three times the posterior width obtained for $f_{mask}$, with a minimum width of 0.02 and a maximum of 0.4 (close to the original prior, \ref{tab:priors}). In this way we keep the nuisance parameter approach for $f_{mask}$, but at the same time we model the improvements that a neural network would learn if $f_{mask}$ was inferred. We don't find any significant difference in the performance compared to the case where the masking fraction is kept fixed to its prior range. 

In the sequential rounds we keep the same NN structure and training setup as in round-0; we also keep the same number of training samples ($N_{samp} = 5 \times 10^5$) in such a way that the density $D$ in the sampling is increased at every round. We train for a maximum of 6 rounds after round-0, typically the entire training of 6 rounds takes less than a day on a GPU. We perform the sequential SBI algorithm on both the composite network and azimuthal network, choosing a reference EMRI with the following parameters: $\theta = \{m_1, m_2, a,e_f,T_{PL}\} = \{10^6 M_{\odot},20 M_{\odot}, 0.7, 0.1, 1.9 \ yr\}$, and three different masking fractions: $f_{mask} = \{ 0.0, 0.15, 0.37\}$.
\subsection{Results}\label{sec:PE_results}

\begin{figure*}
    \includegraphics[width=\linewidth]{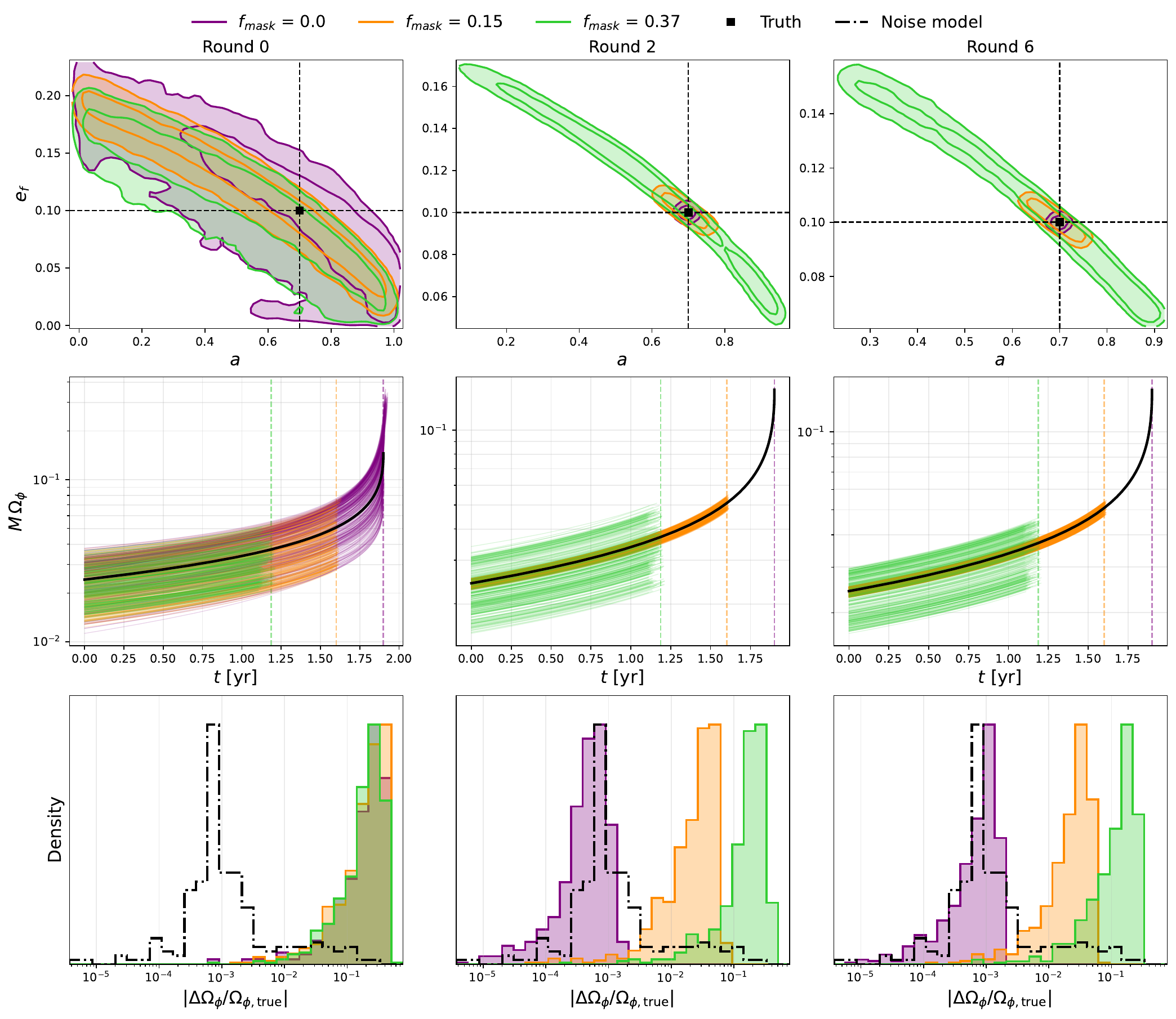}
    \caption{In the top row, the 2D posterior for $a$ and $e_f$ parameters for three different levels of masking $f_{mask} = \{0.0, 0.15, 0.37\}$, coloured in purple, orange and green respectively. On the middle row, the dimensionless obtained from parameter sets $\theta_i = \{m_{1,i}, m_{2, i}, a_i, e_{f,i}, T_{PL, i}\}$ drawn from the full posterior. The vertical lines in the bottom plot represents the truncation of the track due to the different levels of masking fraction. The last row represents the median relative difference of the fundamental frequency tracks inferred by the neural network across the trajectory. The black histogram represents the difference in the same quantity, computed over random realisations of the noise model described in Sec. \ref{sec:dataset}}
    \label{fig:SBI_shrinking_posterior}
\end{figure*}

To illustrate the shrinking process of the posterior distribution discussed in the previous section, we choose the final eccentricity ($e_f$) as a reference parameter and display the distribution obtained through round-0, round-2 and round-6 of the sequential SBI algorithm for three reference masking fractions in the top row of Fig. \ref{fig:SBI_shrinking_posterior}. As can be seen, the marginal posterior distribution shrinks significantly between round-0 and round-2, while the shrinking is more modest between round-2 and round-6, however it gets reshaped for the $f_{mask} = 0.37$ case. In the central row of Fig. \ref{fig:SBI_shrinking_posterior} we illustrate the learning process of the sequential algorithm in track space for the dimensionless $\Omega_{\phi}$. We draw 300 sets of parameters $\theta_i$ from the posterior distribution and compute the relative frequency evolutions through $\tt few$. The tracks for each masking fraction are represented up to the point at which the true track is masked at the injection. As can be seen from the figure, successive rounds of inference focus the track band closer and closer to the true track, following the posterior shrinking. In the last row of Fig. \ref{fig:SBI_shrinking_posterior} we illustrate the shrinking process adopting the median of the absolute relative difference of the tracks sampled from the posterior with respect to the true one. The black histogram represents the same quantity computed on the search outcome from \citep{Speri_2026}, also adopted in the noise modelling as explained in Sec. \ref{sec:dataset}. 

The results shown in Fig. \ref{fig:SBI_shrinking_posterior} demonstrate that the larger the masking fraction, the broader the per-round posterior (after round-0). The reason for the observed behaviour is the distribution of the information across the track: the information to achieve an $O(1) \ \%$ level of precision in the estimates of some of the intrinsic parameters of the EMRI ($e_f, a, m_2$) is contained in the plunging fraction of the inspiral, where the frequency evolution is fast. For this reason if the track is masked in that section the amount of physical information we are able to extract with the SBI process will reflect this limitation. The primary mass $m_1$ and the $T_{PL}$ are easier to estimate with high precision because they set the frequency scale and the duration of the signal, which are simpler features to learn with the neural networks. Moreover, as can be seen from Fig.~\ref{fig:SBI_shrinking_posterior}, the level of accuracy obtained in the $f_{mask} = 0$ case is comparable to the noise injected, reaching the maximum precision allowed by the search pipeline. It is not the same for higher masking fractions, for which the truncation for the track does not allow for the maximum level of shrinkage.

\begin{figure}
    \includegraphics[width=\linewidth]{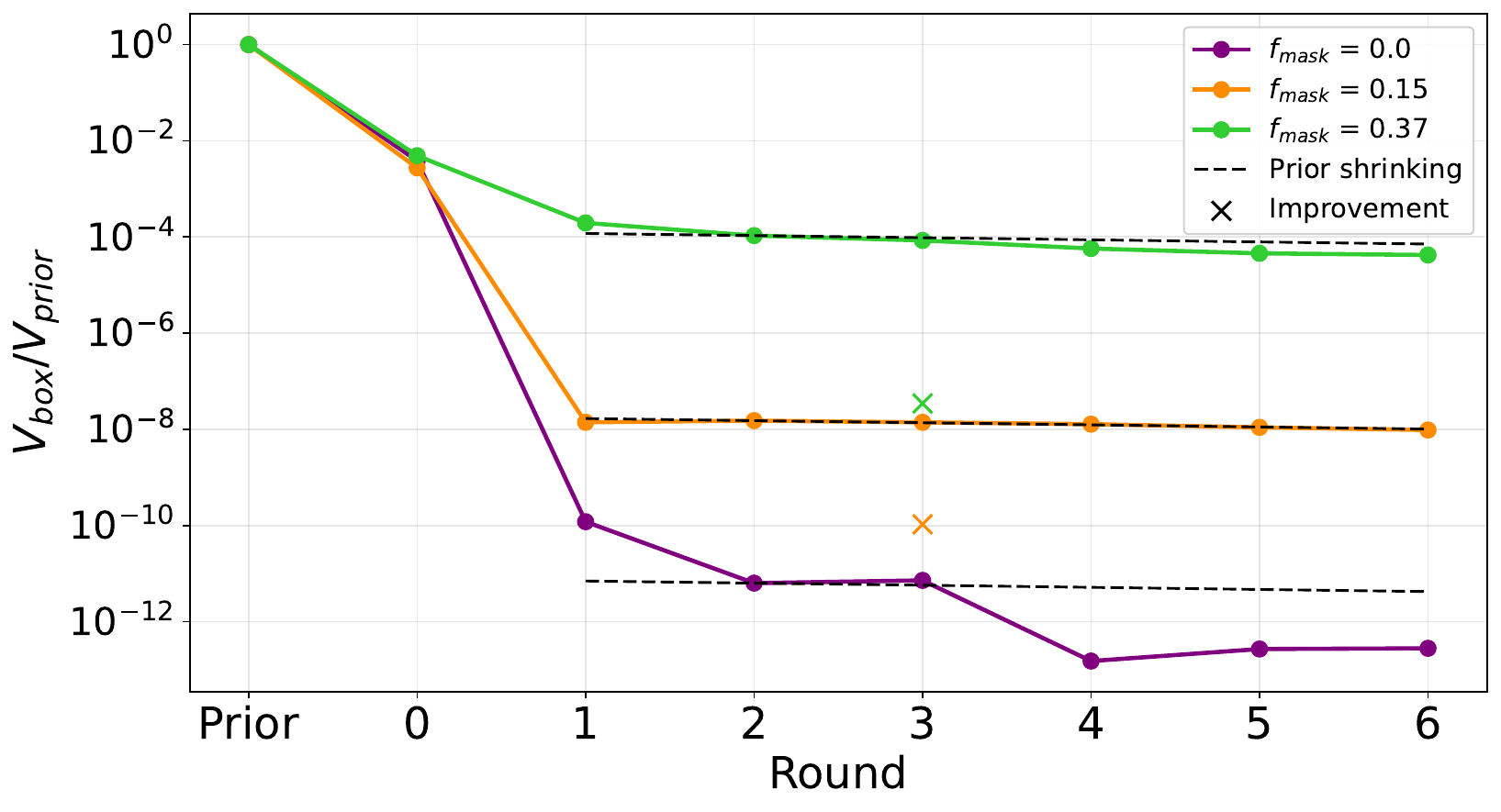}
    \caption{The shrinking of the prior box volume (defined in Sec. \ref{sec:PE}) across every round of inference, normalized to the astrophysical prior volume, for three different levels of masking: $f_{mask} = \{0.0, 0.15, 0.37\}$, coloured in green, blue and red respectively. The dotted lines represent the shrinkage of the prior box volume due to the cut of the new box, chosen to be the 1 \% - 99 \% quantiles at every round. The green and orange crosses represent the improvement in the constrained prior volume applied to the detection statistics discussed in Sec. \ref{sec:PE_improvements} on the posterior samples}
    \label{fig:Prior_shrinking}
\end{figure}

In Fig. \ref{fig:Prior_shrinking} we illustrate the shrinking of the prior box volume at every round with respect to the astrophysical prior box displayed in Tab. \ref{tab:priors}. 
The figure shows that after two rounds of inference the only shrinkage of the prior volume in subsequent rounds comes from the $2\%$ reduction in the prior volume induced from clipping of the prior box (as discussed in Sec. \ref{sec:PE_methods}). The box shrinkage due to prior clipping is illustrated in Fig. \ref{fig:Prior_shrinking} by the dotted lines. Each point on the line has been computed adopting the following formula for a given masking fraction and round $i$:
\begin{equation}
    V_{box, i} = V_{box, i-1} \ (0.98)^{N_{pars}}.
\end{equation}
Where $V_{box, i}$ is the volume of prior box at round $i$, and $N_{pars}$ is the number of parameters. The 0.98 factor depends on the quantiles chosen to clip the prior box. A possible consequence of the subsequent shrinkage is the exclusion of the true values of the parameters from the next-round prior volume, leading to biased estimates of the constrained regions. Consequently, we observe that if we stop the sequential algorithm at round-3 (or round-4 for $f_{mask} = 0.0$), we can provide a significant shrinkage of the parameter space, retaining the true values within the posterior range and reducing the computational time of the sequential process. We will discuss these points more in detail in Sec. \ref{sec:PE_improvements}. We highlight that in the $f_{mask = 0.0}$ case, in the training of round-4 the neural network is able to find a sharp peak in the constrained volume of parameter space of the primary and secondary mass, which is the main factor contributing to the drop in $V_{box}/V_{prior}$ at that point on the purple curve in Fig. \ref{fig:Prior_shrinking}. This result suggests that increasing the number of samples drawn from the posterior for training could help to unveil further features and improve the results. However, we will not discuss this possibility further in this work.

As stated in Sec. \ref{sec:PE_methods}, we have also explored the same sequential SBI approach in the case of single frequency networks, in particular we focused on the azimuthal one with an intermediate masking fraction $f_{mask} = 0.15$. We found that the shrinking of the posterior through sequential SBI rounds is significantly reduced in that case: the minimum $V_{box}/V_{prior} \simeq 10^{-3}$ for the single frequency pipeline, roughly six orders of magnitude larger than the results obtained using the composite sequential SBI pipeline.

\subsection{Improving the constraints}\label{sec:PE_improvements}
The results presented in the previous section can already inform a parameter estimation pipeline about the interesting region of parameter space to explore. However, due to the extreme sharpness of the EMRI posterior in the parameter space, the box identified has to be further reduced so that it does not contain secondary maxima at which an MCMC-based parameter estimation pipeline could get stuck~\citep{Gair:2008zc,Babak:2009ua,Chua_2022}. We follow the approach illustrated in \cite{Speri_2026}. We build a mock time--frequency representation of the data over which a search pipeline would be applied. We generate our reference source at SNR = 30 using the $\tt few$ python package, then we apply the LISA response function using $\tt fastlisaresponse$  \citep{Fastlisaresponse} to obtain the Time-Delay Interferometry (TDI) variables, and we add Gaussian, stationary noise generated assuming a LISA TDI A channel sensitivity curve. For further details we refer the reader to Sec. F in \cite{Speri_2026}. Then, we draw samples $\theta_i$ from the SBI-shrunk posterior, generate the fundamental frequencies associated with $\Omega_{\phi}(t)$ and $\Omega_r(t)$ and evaluate the detection statistic from \cite{Speri_2026}:
\begin{equation}
    \tilde{S}(\theta) = \sum_{\alpha} \frac{(\mathrm{max}_{\phi_{\alpha}}\langle d_{\alpha}|h_{\alpha}\rangle)^2}{\langle h_{\alpha} | h_{\alpha}\rangle},\label{eq:detstat}
\end{equation}
in which $\alpha$ labels the chunks of the time--frequency representation, $d_{\alpha}$ and $h_{\alpha}$ are the noisy data and the waveform template in the corresponding chunk, $\phi_{\alpha}$ is the waveform phase and $\langle-\rangle$ is the inner product. the maximization is defined as:
\begin{equation}
    \mathrm{max}_{\phi_{\alpha}}\langle d_{\alpha}|h_{\alpha}\rangle = 4 \ \left| \sum_{f} \frac{d_{\alpha}^{*}(f)\ h_{\alpha}(f)}{S_n(f)} \Delta f\right|,
\end{equation}
where $S_n(f)$ is the LISA noise power spectral density (PSD).

\begin{figure*}
    \includegraphics[width=\linewidth]{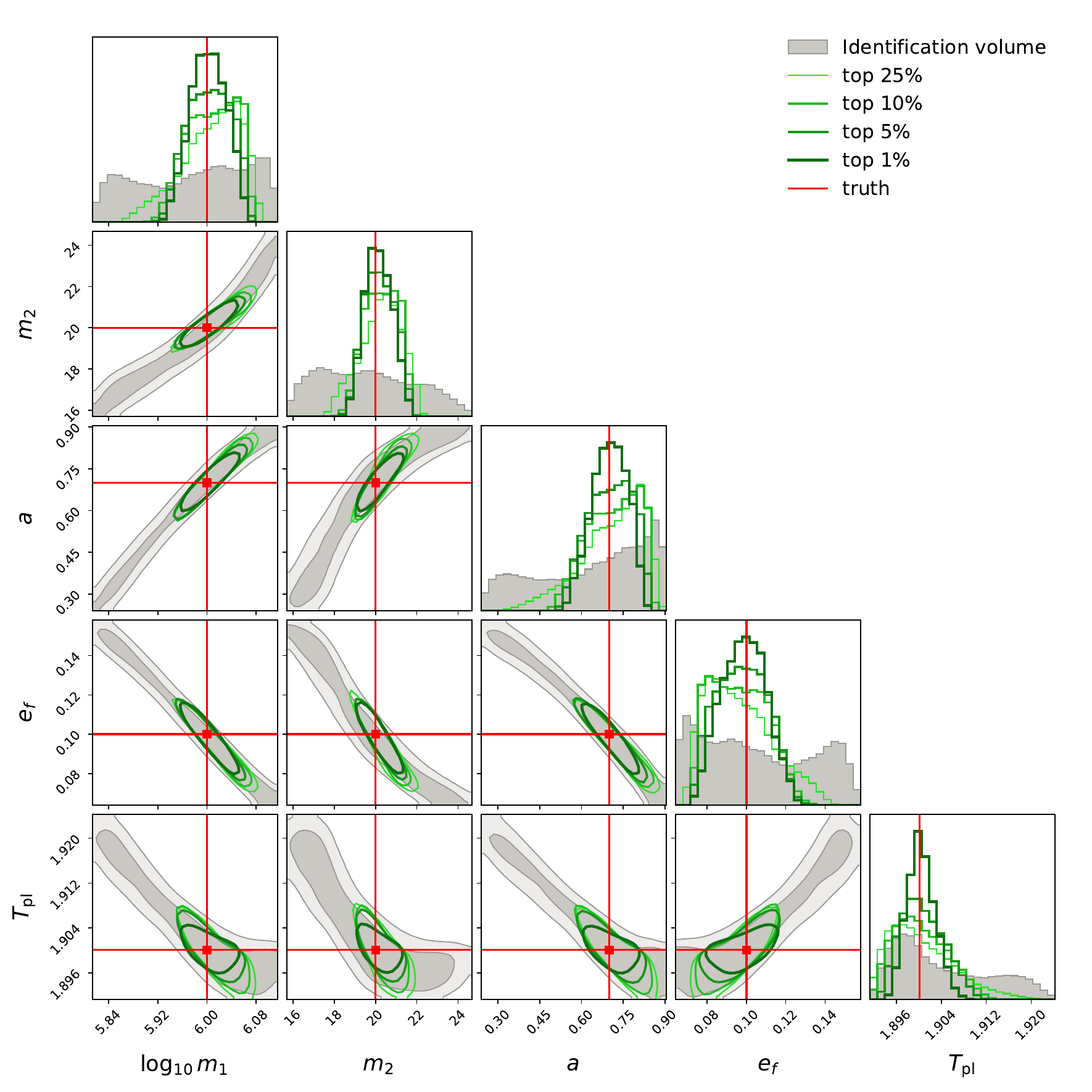}
    \caption{The corner plot of the round-3 posterior obtained by the combined NN in the $f_{mask} = 0.37$ case. We evaluated the detection statistic across $10^6$ random samples from grey distribution and the coloured distributions represent the best detection statistic contours, increasing in detection statistic value from light green to dark green. The red lines represent the injection parameters.}
    \label{fig:corner_cuts_mask037}
\end{figure*}

We evaluate the detection statistics for the $(m = 2, n= 0)$ mode of the signal and select the samples among our draws with the highest detection statistic score. We apply the procedure described starting from the posterior of the composite neural network for the masking level $f_{mask} = 0.37$, which are by far the widest in parameter space. We evaluate the statistic on the spectrogram up to the last visible time imposed by the masking fraction, for instance in this case $T_{vis} \simeq 1.2 \ yr$. As previously stated the prior volume shrinkage stops between round-2 and round-3, so we select the posterior at round-3 as our starting point. We draw $10^6$ samples from the posterior and compute the frequency evolution of the (2, 0) harmonic for each parameter set with $\tt fewtrax$ and evaluate the detection statistic in Eq. \ref{eq:detstat}. The entire procedure takes $O(10)$ minutes on a single A100 GPU. The results are shown in Fig. \ref{fig:corner_cuts_mask037}. 

As observed in the figure, the detection statistic score is a reliable tool to shrink the interesting region of parameter space further than what is obtained through SBI, reducing the prior volume ratio by four orders of magnitude to a final level of $V_{box}/V_{prior} \simeq 3.4 \times 10^{-8}$, which is comparable to the $f_{mask} = 0.15$ case shown in Fig. \ref{fig:Prior_shrinking}. We applied the same procedure to the round-3 posterior of $f_{mask} = 0.15$ case, finding that we are able to achieve a final shrinking factor $V_{box}/V_{prior} \simeq 1 \times 10^{-10}$. We note, however, that the evaluation of the detection statistics in Eq. \ref{eq:detstat} requires selection of a specific harmonic track out of the many that can appear in the signal (see, for example, Fig. 1 in \cite{Chapman_Bird_2025_FastFramDragging}). In the cases where the eccentricity is sufficiently high (typically above $e_f \gtrsim 0.1$), the (2, 0) track may not be the best choice due to the spread of the power contained in the signal to higher harmonics. The consequence of a non-optimal harmonic selection in this procedure is a bias in the constrained region of parameter space due to the presence of the noise in the spectrogram. This limit can be overcome by using the results of Sec. \ref{sec:mode_selection}, where as a follow up of the search pipeline we identify which harmonics were found, which are likely to be among the loudest for the chosen signal. In this section we focussed on the case of the (2, 0) mode of the EMRI. In general, the performance of this strategy will depend on the mode selection procedure: if the harmonic is correctly identified, the detection statistics should be high within the region of parameter space identified by the distribution from the sequential SBI pipeline. However, this could be different for highly eccentric signals, in which the power is spread across many harmonics, and the power-per-harmonic is consequently lower. We have not yet tested the pipeline in that context and leave this to future investigation.

\section{Discussion and Conclusion}\label{sec:DnC}

In this work we explored a possible follow-up strategy to an agnostic search pipeline for EMRIs in time--frequency domain. We summarize our main results, which address the questions raised in Sec. \ref{sec:Intro}. 
\begin{enumerate} 
    \item We demonstrated that given two frequency tracks our pipeline is able to recover the correct harmonic pair associated to both tracks more than $95 \%$ of the time, across parameter space. We explored the main confusion modes in Appendix \ref{sec:Appendix_B}, and proposed possible future improvements to the current pipeline.
    \item We found that assuming a perfect track as search output, the accuracy and precision of the recovery through the sequential SBI process can be brought arbitrarily close to the true values of the %pushed to point estimate level on the 
    intrinsic parameters.
    \item In a more realistic scenario, where the search pipeline recovers noisy tracks, we find that the sequential pipeline is able to marginalize over the noise features and shrink the parameter space volume by $\sim 13$ orders of magnitude with respect to the prior in the best case scenario. If the search output is a masked track (see Sec. \ref{sec:dataset} for more details), the sequential procedure is less effective providing $\sim $ 4 orders of magnitude shrinkage in the worst case scenario. However, we demonstrated that a subsequent application of the same detection statistic used in the search over the region of parameter space proposed by the sequential SBI shrinks the parameter space volume down to a final volume that is $\sim 8$ orders of magnitude smaller than the prior.
\end{enumerate}

Our results underline the importance of having an efficient time--frequency search pipeline able to recover the complete frequency track, because a large part of the physical information content and SNR lie in the non linear fraction of the frequency evolution. Alongside a working search pipeline, it is also important to have a comprehensive knowledge of the limitations and features of the search pipeline across many regimes of SNR and frequency. We also suggest that it is possible to use the mode selection strategy discussed in this work to identify harmonic frequency tracks coming from different EMRI signals present at the same time in the dataset. If among an exhaustive bank of harmonics the procedure can't identify any plausible mode combinations for the frequency track recovered, it may be possible that the frequency tracks are associated with two different systems (see also the discussion in the next section).

\begin{comment}
Provided two frequency evolution tracks, our pipeline can successfully identify the harmonic pair more than $96 \%$ of the times over $2.5 \times 10^4$ different samples of tracks with $T_{PL} = 1.9 \ yr$, ten different masking fractions, intrinsic parameters drawn from astrophysical priors and random noise realizations from a realistic search noise distribution in track space. We explored the cases where the pipeline gets confused in Appendix \ref{sec:Appendix_B}. We demonstrated that a sequential SBI pipeline on frequency evolution tracks alone can shrink the parameter space volume of the intrinsic parameters of $\sim 10^{13}$ orders of magnitude with respect to the astrophysical prior chosen, in few hours of computational time. We also find that in the cases where the shrinking is limited by the fraction of the track available, we can follow up the sequential SBI pipeline with a round of evaluation of the detection statistics applied during the search to reduce the volume of parameter space to a level comparable with the best case scenario quoted above. 
\end{comment}

\subsection{Discussion}
This work provides a first exploration of the space of EMRI time--frequency tracks, investigating the information contained in the frequency evolution of individual harmonics alone. Despite the promising preliminary results, there are several limitations of this work and further improvements are possible. We will divide possible improvements into two categories: \textit{minor} and \textit{significant} changes. We describe the \textit{minor} %changes in the following points:
improvements first:
\begin{itemize}
    \item {\bf Noise model}. As emphasised in Sec. \ref{sec:dataset}, the assumed noise  %characterization 
    model is common for each harmonic frequency track. This may not be true in general and will depend on the reliability of the search pipeline at different frequencies and SNRs. However, after a more detailed analysis of the performance of the search pipeline it will be straightforward to insert the proper noise model into the pipeline: 1) fundamental frequency tracks should be generated; 2) these should be combined under several pairs of harmonics (see point \textit{Harmonic label set} below); 3) the improved noise model should be applied to each harmonic frequency track and then transformed back to fundamental frequencies with Eq. \ref{eq:fund_freq_equations} to obtain the noisy fundamental frequency tracks; and finally 4) train the network using this training data.
    \item {\bf Harmonic labels set}. The harmonic set presented in Sec. \ref{sec:mode_selection_methods} is not exhaustive. In general one wants to try a large number of them, but this procedure is not expensive and requires just one forward pass of each trained network per harmonic label. In our case the noise model adopted was the same for both radial and azimuthal frequency, therefore we can also use the noise independent ratio $Q(t) = \Omega_r(t)/\Omega_{\phi}(t)$ to properly filter realistic harmonic pairs (see Appendix \ref{sec:Appendix_B}). Another option is to train a network to predict the most powerful set of $X$ harmonics given the set of intrinsic parameters as input. $X$ can be any number from 1 to the maximum number of harmonics available for the signal. In this way, we consider only the realistic harmonic set ordered with respect to their power content, which is related to their probability of detection.
    
    \item {\bf Harmonic misclassification}. In Appendix \ref{sec:Appendix_B} we describe the features of the misclassified harmonic sets. In the future, we suggest that our sequential pipeline should be tested on fundamental frequencies coming from a wrong harmonic labelling, and eventually followed up by few MCMC steps to verify the region of parameter space constrained. Moreover, we suggest that another way to improve our pipeline would be to train a simple neural network to predict for a set of intrinsic parameters $\theta$ what are the modes in which most power is concentrated, in this way we would be able to exclude confused but implausible misclassifications.
\end{itemize}
We also propose some \textit{significant} improvements that require more care:
\begin{itemize}
    \item {\bf Different track lengths}. It is possible that the outcome of a search is a set of frequency tracks with clearly different lengths. In this case it may be that the tracks come from two or more different EMRI sources, or that the search pipeline identifies different tracks of the same EMRI with different levels of masking, for example due to a faster frequency evolution of one harmonic compared to the other. In this scenario, we should include this extra ``noise feature'' in the neural network training.
    \item {\bf Phase information}. So far, the pipeline we presented in this work requires no phase consistency between segments. This property makes it general and applicable to different time--frequency search approaches, however it may represent a limitation in terms of constraining power. The tracking of the waveform phase across several months (or years) of signal duration is the key to unlock extremely precise measurements of the intrinsic parameters \citep{EMRI_science_uniquepot, Babak_2017_EMRIscience}. For this reason, a similar behaviour is likely to be observed in this context and so it could be worthwhile to fold phase information into the pipeline, if this is possible.
    \item {\bf Amplitude modulation}. In the same was as the addition of phase information, we can add waveform amplitude information along the track, weighted by the LISA PSD. The amplitude for every harmonic can be computed numerically from $\tt few$ and this information can be added into the neural network training alongside the fundamental frequency evolutions.
\end{itemize}

\section{Acknowledgments}
We would like to thank Stephen Green, Diganta Bandopadhyay, Xue-Ting Zhang, Ludovica Carbone, Philippa Cole and Christian Chapman-Bird for useful discussions, comments and suggestions received during the course of this work.

\begin{appendix} %First appendix
\section{P-P plot diagnostic}\label{sec:Appendix_A}
In this section we illustrate the P-P plots for the azimuthal, radial and composite neural networks developed in Sec. \ref{sec:mode_selection}. 
\begin{figure*}
    \includegraphics[width=\linewidth]{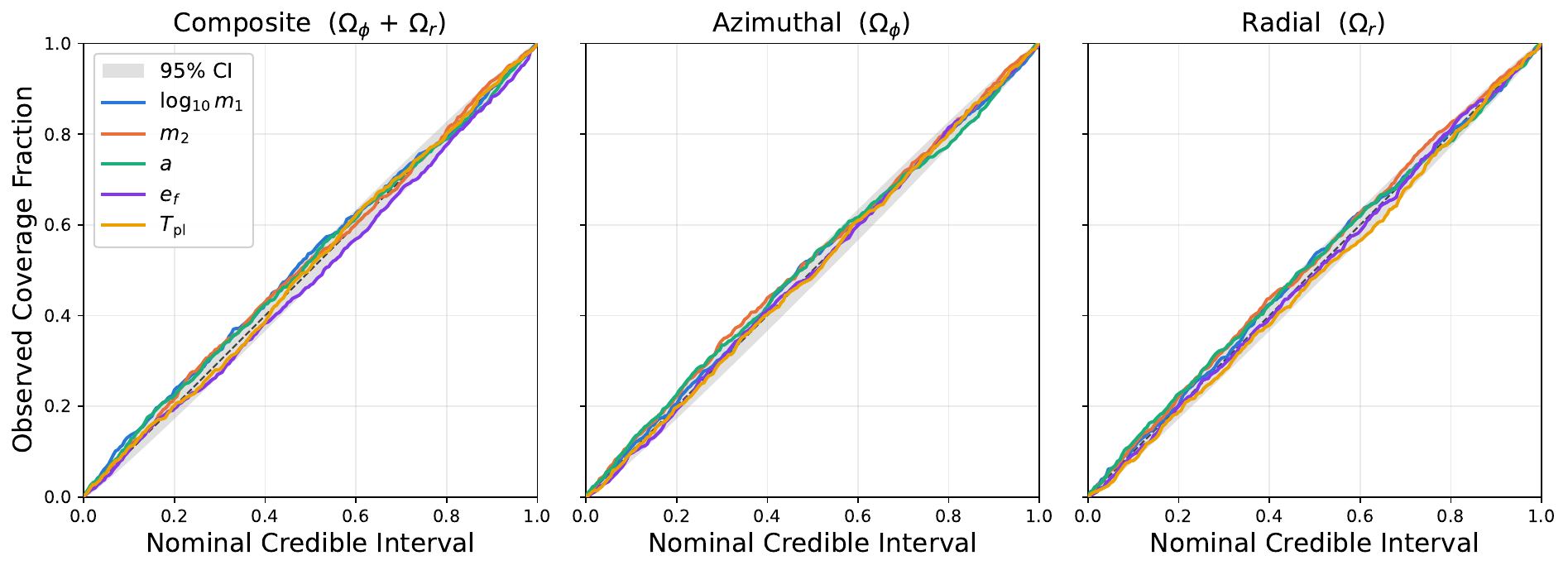}
    \caption{The P-P plots for the three main round-0 neural networks adopted in this work. From left to right: the composite neural network, the azimuthal network and the radial network.}
    \label{fig:PP_plots}
\end{figure*}
Fig. \ref{fig:PP_plots} shows well-calibrated neural networks across the astrophysical priors in Tab. \ref{tab:priors}. We also illustrate the p-values from a Kolmogorov-Smirnov test on the rank distribution for each parameter against a uniform one $\mathcal{U}(0, 1)$ in Tab. \ref{tab:sbc_ks}. We used $N_{\rm PP}=800$ draws from the astrophysical prior and $4 \times 10^3$ posterior draws.

\begin{table*}[t]
  \centering
  \label{tab:sbc_ks}
  \begin{tabular}{lccccc}
    \hline\hline
    Network & $\log_{10} m_1$ & $m_2$ & $a$ & $e_f$ & $T_{\rm pl}$ \\
    \hline
    Composite ($\Omega_\phi$ + $\Omega_r$) & 0.034 & 0.120 & 0.150 & 0.091 & 0.835 \\
    Azimuthal ($\Omega_\phi$) & 0.225 & 0.091 & 0.218 & 0.915 & 0.942 \\
    Radial ($\Omega_r$) & 0.378 & 0.238 & 0.199 & 0.954 & 0.182 \\
    \hline\hline
  \end{tabular}
    \caption{Kolmogorov--Smirnov $p$-values of the Simulation-Based Calibration rank distribution against $\mathcal{U}(0,1)$, for each parameter and each channel network.}
\end{table*}

\section{Classifier confusion sources}\label{sec:Appendix_B}

\begin{table*}[t]
  \centering
  \caption{Ranking performance table. A rival is a \emph{survivor} when the orbit it implies is physical $Q<1$ and the flow actually fitted it ($\mathrm{acc}\ge 0.02$); nothing about it can be rejected without looking at the posteriors. Chance is $\langle 1/(1+n_{\mathrm{rivals}})\rangle$, averaged per set, so it tracks the size of each restricted set.}
  \label{tab:harmonic_id_survivors}
  \begin{tabular}{lllrrr}
    \toprule
    Pair & Population & Ranker & Rivals/set & Top-1 & Chance \\
    \midrule
    $(2,0)+(2,2)$ & all rivals & Classifier & 22.0 & 1.000 & 0.043 \\
     & all rivals & $Q<1$ & 22.0 & 0.191 & 0.043 \\
     & all rivals & acceptance & 22.0 & 0.925 & 0.043 \\
     & survivors & Classifier & 0.6 & 1.000 & 0.785 \\
     & survivors & acceptance & 0.6 & 0.943 & 0.785 \\
    \midrule
    $(2,0)+(2,1)$ & all rivals & Classifier & 23.3 & 0.980 & 0.041 \\
     & all rivals & $Q<1$ & 23.4 & 0.118 & 0.041 \\
     & all rivals & acceptance & 23.4 & 0.520 & 0.041 \\
     & survivors & Classifier & 2.5 & 0.980 & 0.364 \\
     & survivors & acceptance & 2.5 & 0.532 & 0.364 \\
    \midrule
    $(2,1)+(2,2)$ & all rivals & Classifier & 24.0 & 0.988 & 0.040 \\
     & all rivals & $Q<1$ & 24.0 & 0.094 & 0.040 \\
     & all rivals & acceptance & 24.0 & 0.638 & 0.040 \\
     & survivors & Classifier & 2.8 & 0.988 & 0.312 \\
     & survivors & acceptance & 2.8 & 0.645 & 0.312  \\
    \bottomrule
  \end{tabular}
\end{table*}

In this section we investigate in further detail the mode selection pipeline we are adopting. The plain pipeline takes as an input the medians of each parameter from all the neural network posteriors and the acceptance rate of the combined network. With this information it is possible to train a classifier on this summary set to select the most probable pairs of harmonics among the proposed ones. As mentioned in Sec. \ref{sec:mode_selection_methods}, many of the $\sim 2.5 \times 10^4$ cases may represent unphysical configurations, mostly for two reasons: firstly, in our common noise realization approximation, the ratio between the noisy fundamental frequencies $Q = \Omega_{r, \rm noisy}(t)/\Omega_{\phi, \rm noisy}(t) = \Omega_{r}(t)/\Omega_{\phi}(t)$ is a noise-robust quantity. We pointed out in Sec. \ref{sec:dataset} that the frequencies are affected by the same noise realization. In any realistic Kerr spacetime orbit we find $Q < 1$, so we can use the ratio $Q$ to filter for unphysical orbit configurations before passing them to the neural networks and classifier. The second piece of information we may want to introduce is related to the acceptance: through Eq. \ref{eq:fund_freq_equations}, wrong harmonic pairs will be associated to wrong (or unphysical) fundamental frequencies and when passed through the composite neural network they will produce extremely low acceptance rates. Selecting a threshold for the acceptance rate $\alpha_T = 0.02$ (arbitrarily chosen), we can filter and exclude configurations that produce acceptance rates lower than this threshold. In Tab. \ref{tab:harmonic_id_survivors} we investigate different possible scenarios. For each harmonic pair considered, we fix a population that can be the totality of population ($\sim 2.5 \times 10^4$ candidates), labelled as \textit{all rivals} or the \textit{survivors} population obtained after applying the $Q < 1$ and acceptance filters. We compute the average number of rivals that remain in the simulation set after each filter (Rivals/set column). At this point, we select the strategy to discriminate the true harmonic pairs from the wrong ones (Ranker column); several strategies to extract the true pair of harmonics are considered: the filtering of the physical candidates followed by a random draw among them ($Q < 1$), the filtering followed by the ranking by their acceptance ($acceptance$) and the proper classification. We then compute the probability that the true pair has been classified first by our strategy (Top-1 column). We find that the Classifier outperforms the simpler $Q < 1$ and acceptance strategies across all the harmonic pairs tested, illustrating how a blind application of the classifier strategy can provide reliable results. The table also shows that the physical filtering $Q < 1$ is a powerful tool to reduce the rivals for each set, limiting the effective survival population to a few available cases; however, as can be seen from the Top-1 column, the follow up of a classifier is necessary to reach very high probabilities. These tests also show that our mode pipeline is robust to a blind injection of tracks with an ``all rivals'' population, infact the Top-1 probability between the Classifier on ``all rivals'' and ``survivors'' is almost the same in every case.

We also explored the main sources of confusion of our pipeline, trying to identify what are the critical cases where the combination of neural network and classifier is failing. We identified five \textit{families} of confusing configurations based on the structure of the $2\times 2$ matrix $A = M^{-1}M_{O}$, where $M$ is the matrix connecting the harmonic frequencies to the fundamental frequencies given a pair of harmonic labels $\{(m_a, n_a), (m_b, n_b)\}$. Eq. \ref{eq:track_evol} can be written in a geometric fashion as:
\begin{equation}
\begin{pmatrix}f_a \\f_b \end{pmatrix} \simeq \begin{pmatrix}m_a & n_a \\m_b & n_b\end{pmatrix} \begin{pmatrix}\Omega_{\phi} \\ \Omega_{r} \end{pmatrix}, \label{eq:matrix_notation}
\end{equation}
Where the $2 \times 2$ matrix is $M$. For the true $\{(m_a, n_a), (m_b, n_b)\}$ we label the associated $M$ as $M_O$. It follows from \eqref{eq:matrix_notation} that: $(\Omega_{\phi}, \Omega_r) = M^{-1} (f_a, f_b)$. Therefore, labelling  the correct set of fundamental frequencies as $(\Omega_{\phi, O}, \Omega_{r, O})$, we find:
\begin{equation}
    \begin{pmatrix}\Omega_{\phi} \\ \Omega_{r} \end{pmatrix} = A\begin{pmatrix}\Omega_{\phi, O} \\ \Omega_{r,O} \end{pmatrix},
\end{equation}
Where $A = M^{-1} M_O$. The matrix $A$ encodes the \textit{distortion} of the wrong frequency pair with respect to the original ones. We find five possible configurations of this matrix, leading to different types of misclassification: 
\begin{itemize}
    \item \textit{Rescaling}: $A = \lambda I$, where $\lambda$ is a constant number. The wrong fundamental frequencies are a rescaling of the true ones.
    \item \textit{Diagonal}: $A$ is diagonal, but each fundamental frequency is rescaled independently from the other one.
    \item \textit{$\Omega_{r, \phi}$ Decoupled}: $A$ is upper (lower for $\Omega_{\phi})$ triangular, one frequency is a multiple of the fundamental one, the other is mixed.
    \item \textit{General}: $A$ is a $2 \times 2$ dense matrix and the wrong frequencies are combinations of the true ones.
\end{itemize}
The relative frequency of each of these confusion possibilities depend on the harmonic pair bank we are using. For example, in our case the rescaling appears only in the $\{(2, 0), (2,1)\}$ harmonic pairs, because in the other sets of true pairs tested we do not have a ``rescaled'' harmonic pair set. We also notice that the true harmonic combination $\{(2, 0), (2,2)\}$ is always correctly classified. This may depend on the choice of our harmonic library, which could not be adapt to find competitors to the true pair that pass both the $Q < 1$ and acceptance criteria, as can be seen by the very low Rival/set value in Tab. \ref{tab:harmonic_id_survivors}.
Among the confusion mode families described, we find that, when present, the rescaling is the one that most often leads to confusion, followed by the $\Omega_r$ decoupled and the diagonal. The general and $\Omega_{\phi}$ families do not rise through our investigation. Although our pipeline may mislabel harmonic pairs, we highlight that when the true harmonic is confused the true label appears to be ranked second or third all the times. Concerning the posterior distributions, we observe that in the cases where the true harmonic pair is misclassified, the relative difference between the medians of the posterior associated with the wrong labelling with respect to the medians from true labelling $\mathrm{med}((\theta_{wrong}) - \mathrm{med}(\theta_{true}))/\mathrm{med}(\theta_{true})$ are typically 10 to 100 times larger for the harmonics that are clearly classified as wrong, with respect to the ones that are wrong but missclassified above the truth.

These results suggest that even if the mode selection pipeline misclassifies the harmonic pair, the region of parameter space proposed by the round-0 neural network is the same as the one proposed for a correctly classified harmonic pair. We did not try to run the sequential pipeline in the case of a wrongly identified harmonic pair to check if this parameter space consistency is maintained when the parameter space volume is shrunk through sequential SBI. However, we underline that if the pipeline points to the wrong region of parameter space, a few iterations of an MCMC pipeline should identify that the region of parameter space constrained is not plausible.

\begin{comment}
Secondly, if the ranking of the true harmonic pair set is lower than the third position, the reason is that the posteriors from the neural networks are similar for the different fundamental frequencies and the classifier is not able to rank them correctly. In particular, we find that in the cases where the true harmonic pair is heavily misclassified, the standard deviation (spread) of the medians of the neural network posteriors in each parameter is typically lower with respect to the same standard deviation computed on cases where the true harmonic pair is slightly misclassified, meaning that it is the particular physical parameters of the heavily misclassified configuration that makes the classification difficult. A similar trend can be observed on a larger scale between cases where the true harmonic pair is classified first and the ones where it gets misclassified. We find that the standard deviations of the medians from the composite network is, in general, 2 to 100 times larger on correctly classified cases with respect to the same standard deviation computed over cases where the true pair is misclassified. At the same time, the relative difference of the medians with respect to the medians of the true labelling $\mathrm{med}((\theta_{wrong}) - \mathrm{med}(\theta_{true}))/\mathrm{med}(\theta_{true})$ are typically two order of magnitude larger in the correctly classified cases with respect to the misclassified ones. 
\end{comment}

\end{appendix}

\bibliographystyle{aa}
\bibliography{Refs}
\end{document}